\documentclass[11pt,twoside]{article}

\usepackage{asp2004}
\usepackage{epsf}
\usepackage{graphics}
\usepackage{graphicx}
\usepackage{lscape}

\newcommand{\msun}{\ensuremath{\mathit{M}_{\odot}}}                  
\newcommand{\lsun}{\ensuremath{\mathit{L}_{\odot}}}                  
\newcommand{\zsun}{\ensuremath{\mathit{Z}_{\odot}}}                  

\newcommand{\mdot}{\ensuremath{\dot{M}}}                             

\begin{document}
\title{A Conceptual Analysis of Mass Loss and Frequency Redistribution in Wolf-Rayet Winds}    
\author{Andrew J. Onifer}   
\affil{Los Alamos National Laboratory, X-2, MS-T087, Los Alamos, NM 87545}
\author{Kenneth G. Gayley}
\affil{Department of Physics and Astronomy, University of Iowa, Iowa
City, IA 5 2242}

\begin{abstract} 
To better understand Wolf-Rayet stars as progenitors of gamma-ray bursts, 
an understanding of the effect metallicity has on Wolf-Rayet mass loss is needed. Using simple 
analytic models, we study the $\mdot - Z$ relation of a WN star
and compare the results to similar models.  We find that \mdot~roughly follows a power
law in $Z$ with index 0.88 from $-2.5 \leq \mathrm{log} Z / \zsun \leq -1$ and appears to flatten by
$\mathrm{log} Z / \zsun \sim -0.5$.

\end{abstract}



\section{Introduction}
As a complement to detailed simulations of Wolf-Rayet (WR) winds 
\citep[e.g.,][]{hilliermiller1998, grafenerhamann2005}, we have developed a set of simple analytic models
using diffusive CAK-type line driving with frequency redistribution 
\citep{onifergayley2006}.  In this paper, we apply these models to a preliminary study of the
$\mdot - Z$ relation for WR stars, which is important for the study of WR stars as progenitors
of long-duration gamma-ray bursts \citep{galamaetal1998,iwamotoetal1998,woosleyheger2006}.

\section{Model and Results}
\begin{figure}[!ht]
\includegraphics{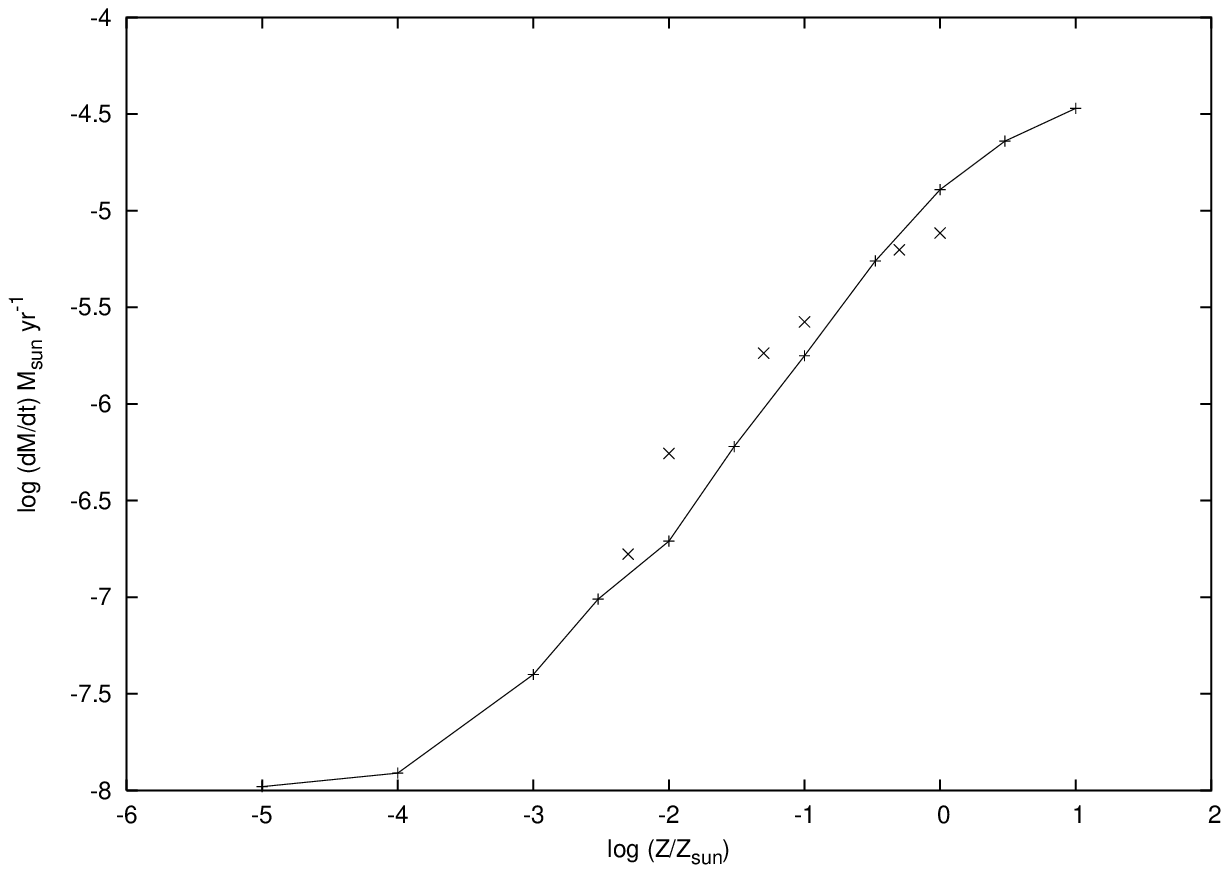}
\caption{\protect\label{vdfig} Log \mdot~vs.~log $Z/\zsun$ for a typical WNL star ($M = 20 \msun,
\mathrm{log}~L/\lsun = 5.62, X = 0.15$).  Plusses and lines are from \citet{vinkdekoter2005}, figure 2.  Crosses are
from this paper.}
\end{figure}

The details of the basic model are discussed in \citet{onifergayley2006}.
We compare
our results with those from \citet{vinkdekoter2005} (hereafter VdK) using their WN star parameters
(see figure \ref{vdfig}). The 
metal abundances were provided by A. Heger (private communication) from a 25 \msun~evolution 
model based on solar abundances from \citet{lodders2003}.  While the $Z$ from these abundances is 
smaller than the traditional $Z \approx 0.02$, uncertainties in other model parameters are expected to
dominate over uncertainties in abundance.

Figure \ref{vdfig} shows the metallicity dependence of our WNL model, as compared to the model
in VdK.  A least-squares fit to our result over $-2.5 \leq \mathrm{log} Z / \zsun \leq -1$ gives 
$\mdot(Z) \propto Z^m$, where $m = 0.88$, very close to the VdK result of $m = 0.86$.  It appears that our 
\mdot~flattens more quickly than those of VdK, perhaps because Fe saturates more quickly with our more
complete line list.  Calculations to higher $Z$ need to be done to confirm that there is indeed 
a flattening, however.

\section{Conclusions}
We have applied analytic WR wind models to the study of \mdot~as a function of $Z$.
Our preliminary results for a WNL-type star show a power-law dependence of \mdot~with $Z$ over 
$-2.5 \leq \mathrm{log} Z/\zsun \leq -1$,
with index $m = 0.88$, similar to the the findings VdK.  However, our \mdot~seems to flatten more quickly.
We plan to cover a wider range in $Z$ to confirm this and to perform the same analysis for
WC stars.

\acknowledgements A. Onifer would like to thank Alexander Heger for helpful discussions
and valuable data. 
Portions of this work were performed under the auspices of the U.S. Department of Energy
by Los Alamos National Laboratory under contract No. W-7405-ENG-36.



\begin{thebibliography}{}

\bibitem[Galama et al.(1998)]{galamaetal1998} Galama, T.~J., et al.\ 
1998, \nat, 395, 670 


\bibitem[Gr{\"a}fener \& Hamann(2005)]{grafenerhamann2005} Gr{\"a}fener, 
G., \& Hamann, W.-R.\ 2005, \aap, 432, 633 

\bibitem[Hillier \& Miller(1998)]{hilliermiller1998} Hillier, D.~J., \& 
Miller, D.~L.\ 1998, \apj, 496, 407

\bibitem[Iwamoto et al.(1998)]{iwamotoetal1998} Iwamoto, K., et al.\ 
1998, \nat, 395, 672 

\bibitem[Lodders(2003)]{lodders2003} Lodders, K.\ 2003, \apj, 591, 
1220 

\bibitem[Onifer \& Gayley(2006)]{onifergayley2006} Onifer, A.~J., \& 
Gayley, K.~G.\ 2006, \apj, 636, 1054 

\bibitem[Vink \& de Koter(2005)]{vinkdekoter2005} Vink, J.~S., \& de 
Koter, A.\ 2005, \aap, 442, 587

\bibitem[Woosley \& Heger(2006)]{woosleyheger2006} Woosley, S.~E., \& 
Heger, A.\ 2006, \apj, 637, 914 

\end{thebibliography}
\end{document}